\documentclass[
    ,final            
  ]
  {aipproc}

\layoutstyle{6x9}
\usepackage{amsmath}
\newcommand{\be}  { \begin{equation} }
\newcommand{\ee}  { \end{equation}   }
\newcommand{\bea} { \begin{eqnarray} }
\newcommand{\eea} { \end{eqnarray}   }

\newcommand{\msun}{ \ensuremath{M_{\odot}  } }

\newcommand{\Phic}{ \ensuremath{\Phi_\mathrm{c}} }
\newcommand{\apj}{ApJ~}
\newcommand{\apjl}{ApJL~}
\newcommand{\aap}{A\&A~}
\newcommand{\mnras}{MNRAS~}
\newcommand{\nat}{Nature~}

\newcommand{\Apc}{\ensuremath{A_\mathrm{pc}}}

\begin{document}

\title{PSR B1257+12: a quark star with planets?}


\author{Youling Yue}{
  address={Astronomy Department, School of Physics, Peking University,
           Beijing 100871, China}
}

\author{Renxin Xu}{
  address={Astronomy Department, School of Physics, Peking University,
           Beijing 100871, China}
}

\begin{abstract}
A recent observation has shown that PSR B1257+12 could have  quite
small X-ray emitting area, only about 2000 m$^2$, which is more
than three orders smaller than the canonical  polar cap size.
We suggest here that PSR B1257+12 could be a low-mass quark star
with radius of $R \simeq 0.6$ km and mass of $M \simeq
3\times10^{-4}\msun$.
Such a low-mass quark star system may form in an accretion induced
collapse process or a collision process of two quark stars.
\end{abstract}

\maketitle

\section{Introduction}

The first three extrasolar planets
\cite{Wolszczan1992Natur1257discover} was discovered around a
millisecond pulsar, PSR B1257+12, with spin period of $P = 6$ ms.
Dispersion measure shows a distance of about 500 pc to pulsar, and
X-ray spectrum reveals a hydrogen column density of $N_{\rm H} =
3\times10^{20}~\mathrm{cm^{2}}$ \cite{Pavlov2007ApJ_1257_xray}.
The parameters of the three planets are listed in Table 1.
\begin{table}[htbp]
\caption{Parameters of PSR B1257+12's planets: orbital period
$P_{\rm orb}$, mass $m_{\rm p}$, orbital radii $a_{\rm p}$
\cite{KonackiWolszczan2003_1257pl_mass_angle}, mass from this
work $m_{\rm p,new}$, and orbital radii from this work $a_{\rm
p,new}$.} \label{tab:1}
\begin{tabular}{ccccccc}
\hline
Planet
& $P_{\rm orb}$ (day)
&$m_{\rm p}$ ($M_{\rm earth}$) \tablenote{assuming a central pulsar of $M=1.4\msun$.}
&$a_{\rm p}$ (AU) $^*$
& $m_{\rm p,new}$ ($M_{\rm earth}$)
\tablenote{this work, assuming a central pulsar of
$M=3\times10^{-4}\msun$, see text.}
&$a_{\rm p,new}$ (AU) $^\dag$\\
\hline
A &25.3&0.02&0.19&0.000072&0.011\\
B &66.5&4.3 &0.36&0.016  &0.022\\
C &98.2&3.9 &0.46&0.014 &0.027\\
\hline
\end{tabular}
\end{table}
A recent observation has, however, shown that the millisecond
pulsar B1257+12 could have very small X-ray emitting area, only
about 2000 m$^2$ \cite{Pavlov2007ApJ_1257_xray}, which is
remarkably smaller than the expected value assuming a pulsar with
$R \sim 10$ km. The observation covered 20 ks and received 25
photons. Both power law and blackbody models are acceptable. The
blackbody fitting gives: $kT \simeq 0.22$ keV and projected
emitting area $A \simeq 2000$ m$^2$. The bolometric luminosity is
$2L_{\rm bol} \sim 3\times 10^{29} \mathrm{~ergs~s^{-1}}$
\cite{Pavlov2007ApJ_1257_xray}.

For a pulsar with $R=10$ km and $P = 6$ ms, the polar cap area is
$1.1 \times 10^7 \mathrm{~m^2}$, nearly four orders higher than
the fitted area ($A \simeq 2000$ m$^2$).
In this article, we will demonstrate that this discrepancy could
be explained by assuming that PSR B1257+12 is a low-mass quark
star (QS, for a general review, please see~\cite{Xu2007} in this
proceedings).
The precessing radio pulsar, PSR B1828-11, could also be a
low-mass QS torqued by a quark planet orbiting
around~\cite{Liu2007}.
We discuss the possibility of planet formation under this new
scenario, that differs from the models in which PSR B1257+12 is a
normal neutron star of $M=1.4\msun$.

\section{A low-mass quark star model}

For a  pulsar of  $R = 10$ km and $P = 6$ ms,
the theoretical  value of the polar cap area is
$A_{\rm pc,th} \simeq 2\pi^2 R^3 /(cP) = 1.1 \times 10^7 
\mathrm{~m^2}$ (where $c$ is the light speed),
which is $\sim5000$ times larger than the observation.
We consider that PSR B1257+12 is a low-mass
quark star.
Its radius is $R_{\rm new} = 0.6$ km  to fit the
observed emitting area
$A \sim 2000 $ m$^2$.
Hence,
the mass of the quark star is $M_{\rm new} = 3\times10^{-4}\msun$.

Consequently, the planet's parameters such as the
masses and distances to the central pulsar ( according to the
third Keplerian law) would  be changed,
\[
\frac{ m_{\rm p,new} }{ m_{\rm p} } \simeq (\frac{R_{\rm new}}{10 ~{\rm
km}})^2 =(\frac{A}{1.1\times10^7})^{2/3}
, \quad \frac{ a_{\rm p,new} }{ a_{\rm p} } \simeq \frac{R_{\rm new}}{10
~{\rm km}} \simeq (\frac{A}{1.1\times10^7})^{1/3}
.
\]
Through the third Keplerian law, we can obtain that
$a_{\rm p} \propto R$.
However, the orbital radius of the pulsar ($a_{\rm M}$) should not depend on
$R$ (or $A$), so as to have the same time residue of pulse TOA (time of arrival)
as the $M = 1.4\msun$ case. So we have $m_{\rm p}\propto R^2$.

The new values, $m_{\rm p, new}$ and $a_{\rm p, new}$, are
presented in Table 1. We also show in Fig. 1 the the masses
($m_{\rm p}$) and the orbital radii ($a_{\rm p}$) of the planets
as functions of polar cap area ($\Apc$).
\begin{figure}[htbp]
\includegraphics*[width=.85\textwidth]{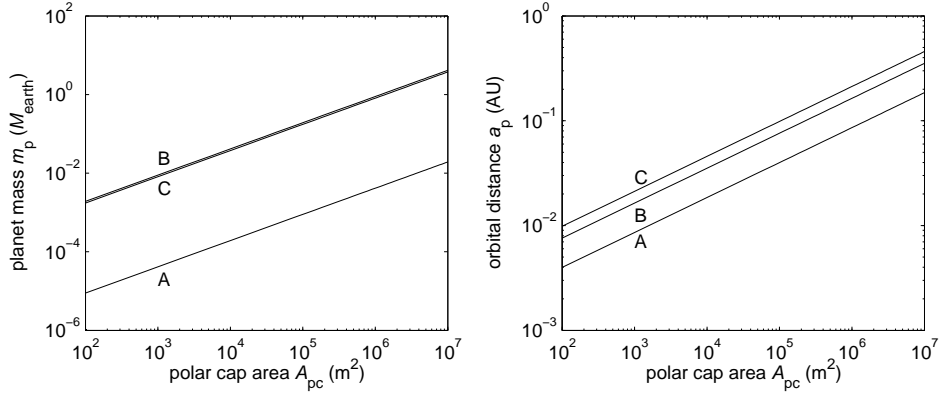}
\caption{The masses and orbital distances of planets as functions of
polar cap area.}
\end{figure}

Small star radius would induce smaller potential drop $\Phi$ in
the polar cap region. The potential drop $\Phi$ should be larger
than a critical value $\Phic$ ($\sim10^{12}$ V) in order to
make a quark star manifest as a radio pulsar
\cite{RS75,UsovMelrose1995}. Considering the effects of
inclination angle \cite{YueCuXu2006}, the potential drop
$\Phi$ in the polar cap region be as larger than $10^{12}$ V if the
inclination angle $\alpha $ is between $10^\circ$ and $60^\circ$.

\section{Discussion}

How the central low-mass quark star form? How the planets form? We
propose two possible scenarios for this pulsar-planets system:
\begin{itemize}
\item 
A low-mass quark star and three quark planets. Such a system may
form after collision of quark stars, and the quark planets could
also be ejected during supernovae explosion of quark star
formation~\cite{Xu2006}.
\item 
A low-mass quark star and three normal planets. Such a system may
form through accretion induced collapse (AIC) process of white
dwarfs \cite{Xu2005} or a quark nova explosion 
\cite{KeranenOuyed03_1257_QN}. 
The planets could probably from in the
fallback disk that forms after the collapse.
\end{itemize}

The low-mass quark star model has advantage for  planet
formation since it has low X-ray luminosity (the disk is cool
and the effect of planet evaporation is thus weak).
The cooling
time scale for the central star is
$
\tau_{\rm cool} \sim V/\mathcal{A} \sim R^3/R^2 \sim R,
$
where $V$ is the stellar volume and $\mathcal{A}$ is the stellar surface area.
A low-mass quark star has
smaller radius and cools more rapidly than a normal neutron star
of $R = 10$ km. So the X-ray luminosity decrease in  a shorter time
scale. This  may make it easier for disk to form and hence
planets formation.

The observation could be fitted by both blackbody and power law
function \cite{Pavlov2007ApJ_1257_xray}. Due to the small number of
photons, it is hard to distinguish these two type of models. We
assume that the emission is blackbody in our calculation. If there
is power law component in the emission, the blackbody part would
be smaller. Thus, the central star should be even smaller.

\section{Summary}

Pulsar 1257+12 could possibly be a low-mass quark star (with three
planets) of $R=0.6$ km if its X-ray luminosity is from the polar
caps. In this case the mass of the planets would be smaller by a
factor of 0.0036. The planets around the low-mass quark star may
form in the fallback disk after an AIC process/a quark nova explosion, through a
collision of quark stars, or as the ejecta during quark star
formation.

\begin{theacknowledgments}
We acknowledge the valuable discussions at the pulsar group of
Peking University. The work is supported by NSFC (10573002,
10778611), the Key Grant Project of Chinese Ministry of Education
(305001), and by the program of the Light in China's Western
Region (LCWR, LHXZ200602).
\end{theacknowledgments}

\end{document}